\documentclass[%
 reprint,
superscriptaddress,
 amsmath,amssymb,
 aps,
]{revtex4-2}

\usepackage{graphicx}
\usepackage{dcolumn}
\usepackage{bm}

\begin{document}

\title{Giant Rydberg excitons in Cu$_{2}$O probed by photoluminescence excitation spectroscopy}

\author{Marijn A. M. Versteegh}
\email{verst@kth.se}
 \affiliation{Department of Applied Physics, KTH Royal Institute of Technology, Stockholm, Sweden}
\author{Stephan Steinhauer}
\affiliation{Department of Applied Physics, KTH Royal Institute of Technology, Stockholm, Sweden}
\author{Josip Bajo}
\affiliation{Department of Applied Physics, KTH Royal Institute of Technology, Stockholm, Sweden}
\author{Thomas Lettner}
\affiliation{Department of Applied Physics, KTH Royal Institute of Technology, Stockholm, Sweden}
\author{Ariadna Soro}
\altaffiliation[Present address: ]{Department of Microtechnology and Nanoscience, Chalmers University of Technology, Gothenburg, Sweden}
\affiliation{Department of Applied Physics, KTH Royal Institute of Technology, Stockholm, Sweden}
\author{Alena Romanova}
\affiliation{Department of Applied Physics, KTH Royal Institute of Technology, Stockholm, Sweden}
\affiliation{Johannes Gutenberg University Mainz, Germany}
\author{Samuel Gyger}
\affiliation{Department of Applied Physics, KTH Royal Institute of Technology, Stockholm, Sweden}
\author{Lucas Schweickert}
\affiliation{Department of Applied Physics, KTH Royal Institute of Technology, Stockholm, Sweden}
\author{Andr\'e Mysyrowicz}
\affiliation{Laboratoire d’Optique Appliqu\'ee, ENSTA, CNRS, \'Ecole Polytechnique,  Palaiseau, France}
\author{Val Zwiller}
\affiliation{Department of Applied Physics, KTH Royal Institute of Technology, Stockholm, Sweden}

\date{\today}

\begin{abstract}
Rydberg excitons are, with their ultrastrong mutual interactions, giant optical nonlinearities, and very high sensitivity to external fields, promising for applications in quantum sensing and nonlinear optics at the single-photon level. To design quantum applications it is necessary to know how Rydberg excitons and other excited states relax to lower-lying exciton states. Here, we present photoluminescence excitation spectroscopy as a method to probe transition probabilities from various excitonic states in cuprous oxide, and we show giant Rydberg excitons at $T=38$ mK with principal quantum numbers up to $n=30$, corresponding to a calculated diameter of 3 $\mu$m. 
\end{abstract}

\maketitle

Rydberg excitons are hydrogen-atom-like bound electron-hole pairs in a semiconductor with principal quantum number $n\geq2$. Rydberg excitons share common features with Rydberg atoms, which are widely studied as building blocks for emerging quantum technologies, such as quantum computing and simulation, quantum sensing, and quantum photonics \cite{adams2020}. Transferring Rydberg physics to the semiconductor environment offers the advantages of the well-developed semiconductor technology, opportunities for integration in microstructures and scalability. In contrast to Rydberg atoms, which are relatively stable, Rydberg excitons have potential for ultrafast applications, since they decay quickly. Furthermore, Rydberg excitons can serve as an interface between semiconductor physics and quantum photonics \cite{assmann2020}. The very strong interactions between Rydberg excitons \cite{waltherprb2018} and resulting giant optical nonlinearities at the single-photon level \cite{walthernatcomm2018} are promising for applications such as single-photon emitters and single-photon transistors. 

While Rydberg excitons have been observed in a few material systems \cite{gross1956, agekyan1977, chernikov2014, he2014, hill2015, wang2020}, cuprous oxide (Cu$_2$O) takes a special position as the semiconductor capable of hosting giant Rydberg excitons with principal quantum numbers up to $n = 25$ \cite{kazimierczuk2014}, and recently even up to $n = 28$ \cite{heckotter2020}. Here, we show yellow $n = 30$ Rydberg excitons, which have an estimated diameter of 3.0 $\mu$m (for comparison, Rydberg atoms typically have diameters in the order of 100 nm) and a Rydberg blockade diameter of 23 $\mu$m \cite{kazimierczuk2014}. In the presence of a dilute electron-hole plasma, the Rydberg exciton diameter can become even larger due to screening effects \cite{heckotterprl2018, walther2020}. Cu$_2$O is special in other ways as well, as it is an abundant nontoxic material of which the electronic band structure allows paraexciton lifetimes in the $\mu$s range \cite{mysyrowicz1979} and provides favorable conditions for Bose-Einstein condensation of excitons \cite{snoke2014}. 

Rydberg excitons in Cu$_2$O were identified and studied in electric \cite{heckotterefield2017, heckotterprb2018} and magnetic fields \cite{assmann2016, heckotterscalinglaws2017, zielinska2017, artyukhin2018, farenbruch2020}. Studies of the absorption spectrum revealed quantum coherences \cite{grunwald2016} and brought an understanding of the interactions of Rydberg excitons with phonons and photons \cite{schweiner2016, schone2017, stolz2018}, dilute electron-hole plasma \cite{heckotterprl2018, walther2020}, and charged impurities \cite{kruger2020}. Photoluminescence (PL) studies showed Rydberg states up to $n = 10$ \cite{takahata2018}, where line broadening could be explained by exciton-phonon and exciton-exciton interactions \cite{kitamura2017}. Phonon scattering from Rydberg states was studied by resonant Raman scattering spectroscopy \cite{yu1978} and transitions from $1s$ yellow excitons to Rydberg states by infrared Lyman spectroscopy \cite{jorger2003, yoshioka2007}. A nonlocal optical response showed up in the PL from Rydberg excitons in Cu$_2$O films with a thickness in the order of the optical wavelength \cite{takahatanonlocal2018}. Recently, theoretical studies were published on the physics of the more highly energetic green exciton series \cite{rommel2020}, as well as on radiative interseries transitions between Rydberg excitons \cite{kruger2019}. An experimental paper presented signatures of coherent propagation of blue exciton polaritons \cite{schmultzler2013}. Electromagnetically induced transparency may be realized for Rydberg excitations \cite{walthernatcomm2018, ziemkiewicz2020} and could enable control over exciton-phonon interactions \cite{walthereit2020}. High-quality synthetic Cu$_2$O microcrystals have been demonstrated as host material for Rydberg excitons \cite{steinhauer2020}, with potential for quantum confinement \cite{konzelmann2020} and applications such as quantum sensing and single-photon emission, where the Rydberg blockade can prevent the excitation of more than one Rydberg exciton at the same time \cite{khazali2017}.

An assessment of the opportunities of Rydberg excitons for quantum technology requires measurement of transitions to and from the Rydberg states. Here, we present photoluminescence excitation (PLE) spectroscopy as a method to explore those transitions. In our approach, a PLE spectrum is constructed by scanning an excitation laser over all excitonic resonances of a Cu$_2$O crystal, except the yellow $1s$ excitons, from which the optical emission is recorded. Thus, the PLE spectrum provides information about the transition probabilities to the $1s$ orthoexciton state. We introduce a comprehensive analytical model to describe our observations. Moreover, we show that PLE spectroscopy allows for the observation of giant Rydberg excitons of the yellow series with principal quantum numbers up to $n=30$. Excitons of the green and blue series are observed as well. In contrast to optical transmission spectroscopy, where one has to work with fragile thin slices of Cu$_2$O crystal (10-100 $\mu$m), PLE spectroscopy is also suitable on samples with large thicknesses or complex structures. 

\begin{figure}
\includegraphics[width=0.47\textwidth]{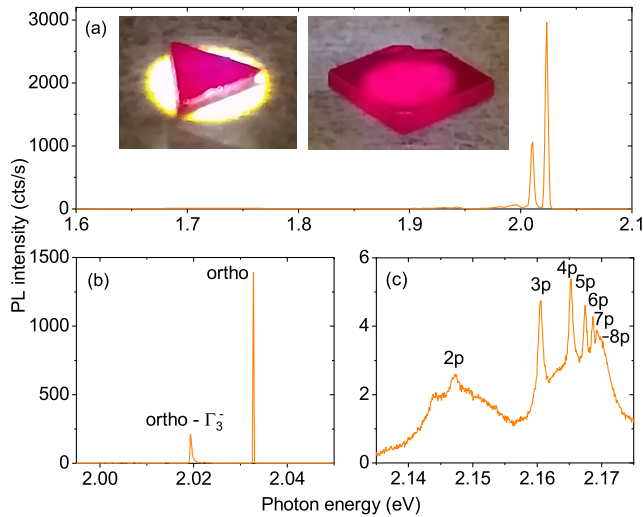}
\caption{\label{fig:PL} Photoluminescence spectra of natural Cu$_2$O. (a) Yellow $1s$ orthoexciton emission and absence of defect emission at 1.7 eV. Excitation conditions: $E_\textrm{exc}=2.17$ eV, $P=1$ $\mu$W, $T=170$ mK. Insets are photographs of the Tsumeb (left) and the Onganja crystal (right). (b) Zoom-in of (a), recorded with a different grating. (c) Yellow Rydberg exciton emission. Excitation conditions: $E_\textrm{exc}=2.30$ eV, $P=80$ $\mu$W, $T=1.0$ K. Presented spectra are of the Tsumeb crystal; spectra of the Onganja crystal are similar.}
\end{figure}

\textit{Methods.} Experiments have been performed on two natural Cu$_2$O (cuprite) crystals [Fig. \ref{fig:PL}(a)]: a flat crystal (triangle with 4 mm sides, 840 $\mu$m thick), originally from the Tsumeb mine in Namibia and part of the same specimen that was used in previous experiments \cite{mysyrowicz1979, snoke1987, snoke1990}, and a (1 0 0) oriented square with dimensions $5\times5\times1$ mm$^3$, purchased from Surface preparation laboratory and originally from the Onganja mine in Namibia. We observed the highest Rydberg states in the Tsumeb crystal, but in general, the two crystals show very similar behavior. Crucial for the appearance of giant Rydberg excitons is the quality of surface preparation (details in the Supplemental Material \cite{supplementalmaterial}). A polished and etched crystal was positioned inside a dilution refrigerator (Bluefors), with windows for optical measurements. The temperature, measured directly next to the crystal, was 38 mK at an excitation power of 2 nW. As excitation laser we used a tunable 1-MHz linewidth CW laser (H\"ubner Photonics C Wave). The laser frequency was stabilized using a HighFinesse wavemeter to within 10 MHz (or 40 neV). The laser power was stabilized to within 0.1\% using a MEMS fiber-optic attenuator (details in the Supplemental Material \cite{supplementalmaterial}). Inside the dilution refrigerator, the excitation laser beam was focused by a lens ($f = 3.1$ mm, NA = 0.70) onto the surface of the crystal, which was mounted on top of a 3-directional Attocube piezoelectric positioner stack, allowing sample alignment with respect to the laser focus with nanoscale precision. The emitted light was collected by the same lens and its spectrum was measured by a Jobin Yvon 55-cm monochromator and an Andor iDus CCD. 

PL spectra are presented in Fig. \ref{fig:PL}. The two dominant emission lines are the direct $1s$ orthoexciton quadrupole emission at 2.033 eV and the $\Gamma_{3}^{-}$-phonon-assisted $1s$ orthoexciton emission at 2.020 eV. The nearly complete absence of broad emission at 1.7 eV (the maximum is 250 times lower than the direct orthoexciton emission maximum) indicates that there are almost no oxygen vacancies within the probed sample volume. For the construction of the PLE spectrum, the excitation laser was scanned with step sizes ranging from 0.2 $\mu$eV in the region with the highest Rydberg states to a few meV in spectral regions far away from any resonances. The total count rate of the two $1s$ orthoexciton emission lines was recorded for each step with integration times between 15 s and 200 s, depending on the chosen excitation power. At the lowest excitation power of 2 nW, a sliding average was calculated of 7 subsequent steps, in order to reduce the noise. 

When the spectrometer was tuned to the range 2.14-2.17 eV, the PL spectrum of the yellow Rydberg excitons, up to $n=8$, was observed [Fig. \ref{fig:PL}(c)]. The low-energy shoulder of the $2p$ peak at 2.144 eV had been observed before and its origin has not yet been clarified \cite{takahata2018}.

\begin{figure}
\includegraphics[width=0.47\textwidth]{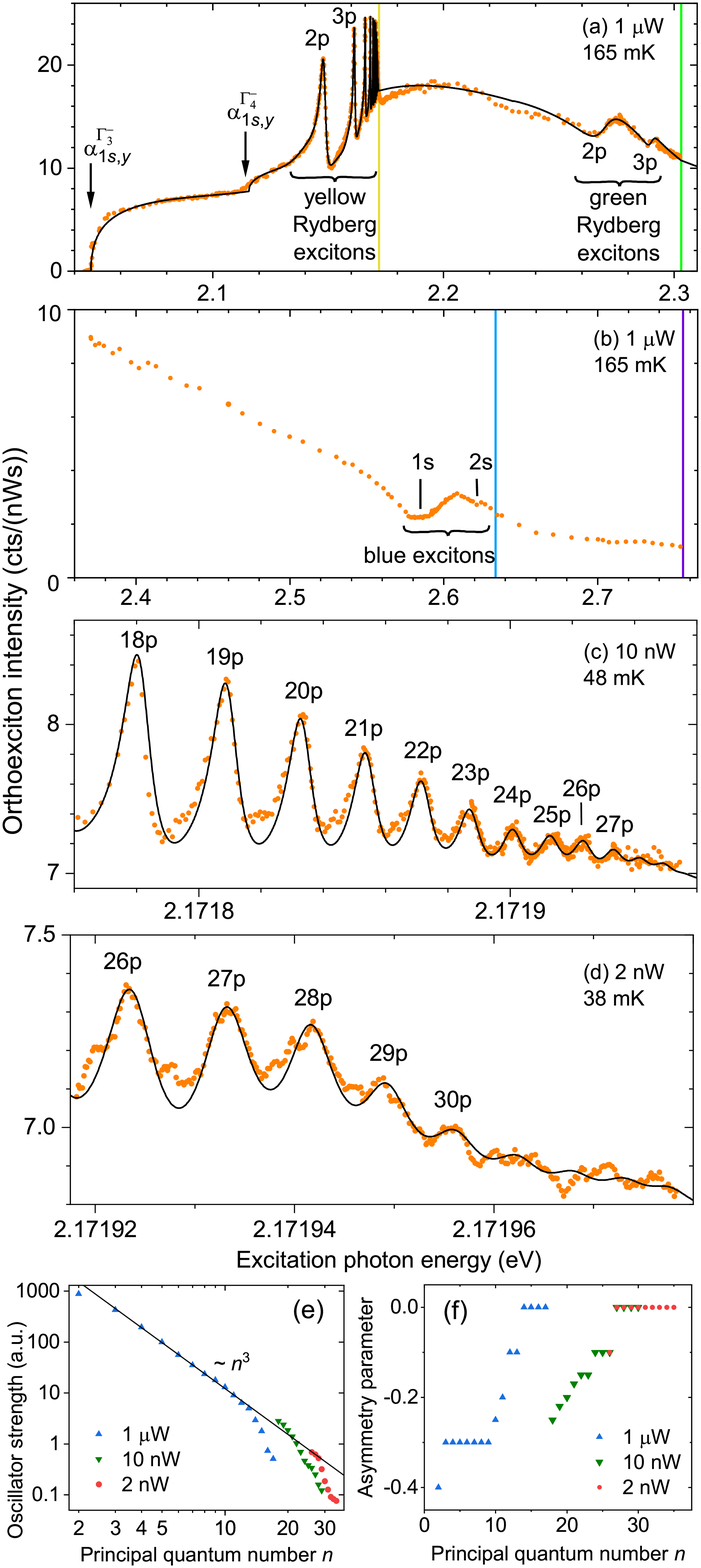}
\caption{\label{fig:PLE} Photoluminescence excitation spectra of natural Cu$_2$O. Detection: $1s$ orthoexcitons. Orange circles: experimental data. Black lines: fits. (a) Yellow Rydberg excitons (positive) and green Rydberg excitons (negative). Vertical lines indicate the yellow and green band gaps. Arrows indicate the start of the $\Gamma_{3}^{-}$ and $\Gamma_{4}^{-}$ phonon-assisted absorption into the yellow $1s$ orthoexciton state. (b) Blue excitons (negative). Vertical lines indicate the blue and indigo band gaps. Giant yellow Rydberg excitons up to (c) $n = 27$ and (d) $n = 30$. Samples are the Onganja crystal for (a) and (b) and the Tsumeb crystal for (c) and (d). (e) Oscillator strengths and (f) asymmetry parameters for the yellow $np$ Rydberg excitons for the indicated excitation powers.}
\end{figure}

\textit{Experimental results.} The measured full PLE spectrum is shown in Figs. \ref{fig:PLE}(a) and \ref{fig:PLE}(b). Scanning the excitation laser from low to high photon energy, the PLE signal starts at 2.047 eV with $\Gamma_{3}^{-}$-phonon-assisted absorption into the yellow $1s$ orthoexciton state (simultaneous creation of an orthoexciton with $E_{1s,y,\textrm{ortho}}=2.0335$ eV and emission of a $\Gamma_{3}^{-}$ phonon with $\hbar\omega_{\Gamma_{3}^{-}}=13.6$ meV) \cite{yu1975}. Around 2.116 eV the $\Gamma_{4}^{-}$-phonon-assisted absorption ($\hbar\omega_{\Gamma_{4}^{-}}=82.1$ meV) into the orthoexcitons starts. In the range of 2.13 eV up to the yellow band gap at 2.172 eV we observe the yellow Rydberg exciton $np$ resonances with well-known asymmetric Lorentzian lineshapes \cite{toyozawa1964}. The strong positive slope of the PLE curve in this region is caused by the start at 2.162 eV of the $\Gamma_{3}^{-}$-phonon-assisted absorption into the green $1s$ orthoexciton state, in combination with the absorption into the Urbach tail of the yellow band gap. At 2.20 eV, above the yellow band gap, a maximum is reached in the PLE signal, where the mentioned phonon-assisted absorption processes are combined with absorption into the yellow continuum states. The maxima of the yellow Rydberg resonances are higher than the PLE signal maximum at 2.20 eV, indicating that yellow Rydberg excitons have a much higher probability of decaying into a yellow $1s$ orthoexciton than other excitations in this energy range, as will be further analyzed in the theory section. At 2.269 eV and 2.290 eV the $2p$ and $3p$ green excitons respectively appear as minima in the PLE signal. The negativity of these peaks indicates that the transition probability from these green exciton states to the yellow $1s$ orthoexciton state is lower than that from other excitations in this energy range. Around 2.584 eV and 2.621 eV we observe the $1s$ and $2s$ blue excitons as negative peaks. Indigo excitons are not visible in the PLE spectrum. The observed spectral positions of the green \cite{rommel2020} and blue \cite{takahata2018} excitons are consistent with literature. 

Zooming in on the yellow Rydberg excitons, we observe the number of Rydberg exciton resonances increasing with decreasing excitation power [Figs. \ref{fig:PLE}(c) and \ref{fig:PLE}(d)], in agreement with literature \cite{heckotterprl2018}.  At 10 nW giant Rydberg excitons up to $n=27$ are visible. At 2 nW Rydberg states up to $n=30$ can be identified, exceeding the highest Rydberg states reported so far \cite{heckotter2020}. As represented by the fitted curves, the measured Rydberg energy levels perfectly follow the Rydberg relation $\hbar\omega_{n}=E_{\textrm{gap}0,y}-\textrm{Ry}_{y}/(n-\delta_{P})^2$ , with the nominal yellow band gap $E_{\textrm{gap}0,y}=2.172053$ eV, the yellow Rydberg constant $\textrm{Ry}_{y}=86$ meV, and the point defect $\delta_{P}=0.21$. The linewidths (full width at half maximum) of the Rydberg excitons are given by the relation $\Gamma_{np,y}=\Gamma_{0}+\Gamma_{1}n^{-3}$, where we find that $\Gamma_{1}=19$ meV and $\Gamma_{0}=5.6$ $\mu$eV at an excitation power of 2 nW and $\Gamma_{0}=7$ $\mu$eV at an excitation power of 10 nW, similar to what was observed with transmission spectroscopy \cite{heckotter2020}. The experimentally determined oscillator strengths, presented in Fig. \ref{fig:PLE}(e), are proportional to $n^{-3}$, but, due to the well-known sensitivity of Rydberg excitons to charges \cite{heckotterprl2018}, as well as to other Rydberg excitons (the Rydberg blockade effect), the oscillator strength for the high Rydberg states is lower, depending on the excitation power. There seems to be absorption into a few Rydberg states with $n>30$ [Fig. \ref{fig:PLE}(d)], and we find that including Rydberg states up to $n=35$ in the theoretical spectrum gives a better overall fit to the experimental data. However, Rydberg states with $n>30$ cannot be clearly identified as the signal-to-noise ratio in this range is low and the observed maxima and minima do not follow the Rydberg relation.  Decreasing the excitation power below 2 nW did not help to distinguish Rydberg states beyond $n=30$. Features observed between the Rydberg $p$ lines may be due to Rydberg excitons with higher orbital angular momentum \cite{heckotter2020, grunwald2016}. 

\begin{figure}
\includegraphics[width=0.47\textwidth]{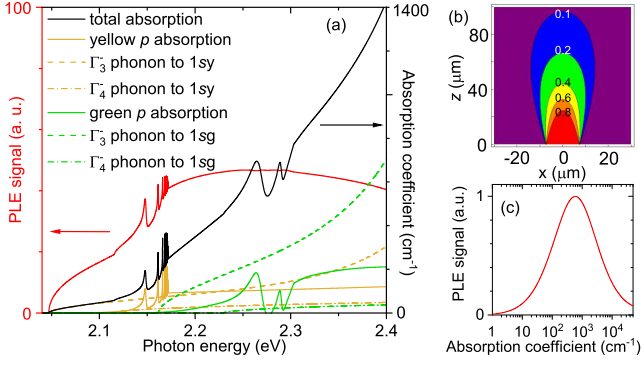}
\caption{\label{fig:theory} Theoretical model for the photoluminescence excitation (PLE) spectrum. (a) Theoretical absorption spectrum (black) and the contributions from the dominant absorption processes (yellow and green lines). Theoretical PLE spectrum according to the model where all absorbed photons create an orthoexciton with the same probability (red). (b) Relative detection probability of a photon emitted at position $(x,0,z)$ in the crystal. (c) Relation between PLE signal and absorption coefficient in the model where all absorbed photons create an orthoexciton with the same probability.}
\end{figure}

\textit{Theory.} In this section, the physical processes underlying the PLE spectroscopy are explained, as well as the shape of the PLE spectrum, which can be fitted according to the black lines in Fig. \ref{fig:PLE}. The absorption spectrum of Cu$_{2}$O in the range of 2.0 to 2.4 eV can be written as the sum of six dominant components \cite{schone2017}:
\begin{equation}
\alpha(\omega)=\alpha_{p,y}+\alpha_{1s,y}^{\Gamma_{3}^{-}}+\alpha_{1s,y}^{\Gamma_{4}^{-}}+\alpha_{p,g}+\alpha_{1s,g}^{\Gamma_{3}^{-}}+\alpha_{1s,g}^{\Gamma_{4}^{-}},
\end{equation}
where $\alpha_{p,y}$ ($\alpha_{p,g}$) denotes the absorption into the yellow (green) $p$ states and $\alpha_{1s,y}^{\Gamma_{i}^{-}}$ ($\alpha_{1s,g}^{\Gamma_{i}^{-}}$) the transition of a photon into a $1s$ yellow (green) orthoexciton with the simultaneous emission of a $\Gamma_{i}^{-}$ phonon. The absorption into yellow $p$ states can be written as
\begin{equation}
\alpha_{p,y}=\alpha_{p,y,\textrm{cont}}+\alpha_{p,y,\textrm{Urbach}}+\sum_{n=2}^{35}\alpha_{np,y},
\end{equation}
where the first term describes the absorption into the continuum above the yellow gap, the second the Urbach tail, and the third the absorption into Rydberg exciton states, with asymmetric Lorentzian lineshapes, where the upper limit of the summation depends on the excitation power. At 2 nW we sum until $n=35$  to get an optimal fit. 
For the absorption into green $p$ states, we have, likewise,
\begin{equation}
\alpha_{p,g}=\alpha_{p,g,\textrm{cont}}+\alpha_{p,g,\textrm{Urbach}}+\sum_{n=2}^{3}\alpha_{np,g}.
\end{equation}
The absorption spectrum with the contribution of the various components is presented in Fig. \ref{fig:theory}(a). More details on the absorption spectrum, including the values for the constants and fitting parameters, are given in the Supplemental Material \cite{supplementalmaterial}.

The relation between the absorption coefficient and the detected PLE intensity is nonmonotonic. The shape of the PLE spectrum can be understood by considering the spatial distribution inside the crystal of the various excitations created by photon absorption, the probability of decay of these excitations into $1s$ orthoexcitons, the diffusion of those orthoexcitons through the crystal and their radiative or nonradiative decay, and finally the probability that a photon emitted by an orthoexciton is collected and detected.

For these experiments, the front crystal surface was positioned carefully in the focal plane of the lens. Recall that the same lens is used for both excitation and collection. From standard Gaussian optics we know that the irradiance inside the crystal of the excitation laser beam, in cylindrical coordinates $r$ and $z$, is, in the presence of absorption $\alpha(\omega)$, given by
\begin{equation}
I_{\textrm{exc}}(r,z,\omega)=\frac{2P}{\pi w_{\textrm{exc}}(z)^{2}}e^{-\frac{2r^2}{w_{\textrm{exc}}(z)^{2}}-\alpha(\omega)z}.
\end{equation}
Here, $P$ is the power of the beam after transmission through the air-crystal interface, $w_{\textrm{exc}}(z)=w_{\textrm{exc},0}\sqrt{(1+(z/z_{R,\textrm{exc}})^2}$ is the beam waist at depth $z$, and $z_{R,\textrm{exc}}=\pi w_{\textrm{exc},0}^2 n_{\textrm{exc}}/\lambda_{\textrm{exc}}$ is the Rayleigh range, with $n_{\textrm{exc}}$ the index of refraction at excitation wavelength $\lambda_{\textrm{exc}}$ and $w_{\textrm{exc},0}$ the beam waist at the focus (at the front crystal surface).
The generation rate of excited states per unit of volume is given by
\begin{equation}
G(r,z,\omega)=\frac{I_{\textrm{exc}}(r,z,\omega)\alpha(\omega)}{\hbar\omega}.
\end{equation}
These excited states relax into orthoexcitons, with a probability depending on their state and energy. As initial model, let us assume that all excited states relax into orthoexcitons with the same probability. Later, we will consider a more refined model about the transition probabilities. 

Orthoexcitons have a lifetime of 3 ns \cite{mysyrowicz1979}, which is large enough to make their diffusion through the crystal non-negligible. Some orthoexcitons recombine nonradiatively at the surface and a small fraction diffuses away from the excitation beam to a part of the crystal from which the emission is not collected. Orthoexciton and paraexciton diffusion have been studied extensively \cite{trauernicht1986, morita2019}. We modeled the orthoexciton diffusion using rate equations as in \cite{yoshioka2010}, with a diffusion parameter of 300 cm$^{2}$/s, and with the condition that if orthoexcitons reach a surface of the crystal (only the front surface is relevant in our case) they vanish due to nonradiative recombination. 

The final element to be considered is the detection probability as a function of the emitting orthoexciton position $(x, y, z)$ in the crystal (the focus is at $(0,0,0)$). This is a standard mathematical problem in microscopy \cite{webb1996,ahi2017}. Details of our calculation are given in the Supplemental Material \cite{supplementalmaterial}. The calculated relative detection probability is shown in Fig. \ref{fig:theory}(b), for the case of $y=0$. We find that the detection probability is almost constant from $z=0$ to $z=20$ $\mu$m, after which it decreases rapidly. 

Taking into account the shape of the excitation beam, the diffusion and surface recombination of the orthoexcitons, and the position-dependent detection probability of emitted photons, we get a relation between absorption coefficient and PLE signal as represented in Fig. \ref{fig:theory}(c). When the optical penetration depth is long, the PLE signal increases linearly with the absorption coefficient, as photons emitted by orthoexcitons close to the focus have a larger probability of getting detected than photons from deep inside the crystal. However, at short penetration depths, the effects of diffusion and surface recombination become dominant, and therefore the PLE signal increases sublinearly at larger absorption coefficients, and decreases for absorption coefficients \mbox{$>600$ $\textrm{cm}^{-1}$}. 

Combining this relation between the PLE signal and the absorption coefficient with the theoretical absorption spectrum, the PLE spectrum is obtained as represented by the red curve in Fig. \ref{fig:theory}(a). Comparing with the experimental data [Fig. \ref{fig:PLE}(a)], it can be concluded that this model does not yet fully describe the relevant physics. For example, in the experimental spectrum the yellow Rydberg excitons have a larger amplitude and a higher maximum than the above-band-gap absorption maximum at 2.20 eV. What therefore needs to be included in the model is a dependence of the transition probability to the thermalized $1s$ orthoexciton state on the type and energy of the photon-generated excited state. Our experimental data suggests the following transition probabilities. For the phonon-assisted absorption into yellow and green $1s$ excitons 
\begin{align} 
P_{1s,y}^{\Gamma_{i}^{-}}(\omega)&=e^{-(\hbar\omega-E_{1s,y,\textrm{ortho}}-\hbar\omega_{\Gamma_{i}^{-}})/E_{\phi}},\\
P_{1s,g}^{\Gamma_{i}^{-}}(\omega)&=\frac{3}{4}de^{-(\hbar\omega-E_{1s,g}-\hbar\omega_{\Gamma_{i}^{-}})/E_{\phi}}.
\end{align} 
For the yellow and green Rydberg excitons
\begin{align} 
P_{np,y}(\omega)&=\frac{3}{4},\\
P_{np,g}(\omega)&=\frac{3}{4}h.
\end{align} 
For the continuum states
\begin{align}
P_{p,y,\textrm{cont}}(\omega)&=\frac{3}{4}de^{-[(\hbar\omega-E_{\textrm{gap},y})/E_{i}]^2},\\
P_{p,g,\textrm{cont}}(\omega)&=\frac{3}{4}hde^{-[(\hbar\omega-E_{\textrm{gap},g})/E_{i}]^2}.
\end{align}
For the Urbach tails
\begin{align}
P_{p,y,\textrm{Urbach}}(\omega)&=\frac{3}{4}d,\\
P_{p,g,\textrm{Urbach}}(\omega)&=\frac{3}{4}hd.
\end{align}
Here, $i=3,4$, $E_{\textrm{gap},y}$ ($E_{\textrm{gap},g}$) is the renormalized yellow (green) band gap, and for the fit parameters we obtain the values $E_{\phi}=0.28$ eV, $d=0.687$, $E_i=187$ meV, and $h=0.187$. Also the energy of the $1s$ green exciton $E_{1s,g}$ is used as a fit parameter. The $1s$ green exciton does not have a single resonance energy, as it is, due to the strong coupling with the yellow $2p$ exciton, spread over several resonances \cite{schweiner2017}. For simplicity, we model a single resonance energy and obtain the best fit with $E_{1s,g}=2.148$ eV, which is equal to $E_{2p,y}$. A physical interpretation of these transition probabilities is discussed in the Supplemental Material \cite{supplementalmaterial}.

The full range of the fit is presented as the black curve in Fig. \ref{fig:PLE}(a). The same fit is used in Figs. \ref{fig:PLE}(c) and \ref{fig:PLE}(d), with oscillator strengths and asymmetry parameters given in Figs. \ref{fig:PLE}(e) and \ref{fig:PLE}(f).  The negative shape of the green Rydberg excitons can be explained by the low probability of $\frac{3}{4}g=0.14$ of reaching a $1s$ yellow orthoexciton state via a green Rydberg excitation, compared to other excitations in the same energy range, such as the phonon-assisted $1s$ yellow and green excitations.

The calculation of the absorption spectrum as presented in Fig. \ref{fig:theory}(a) makes use of second-order perturbation theory \cite{schone2017}, with blue excitons as virtual states, and therefore cannot be accurately extended to energies close to the blue exciton resonances. Qualitatively, the negative shape of the blue excitons [Fig. \ref{fig:PLE}(b)] can be understood from the high absorption coefficient at the $1s$ and $2s$ blue exciton resonances (dipole-allowed transitions) \cite{daunois1966}, combined with the negative slope of the PLE-signal versus absorption coefficient relation at high absorption coefficients [Fig. \ref{fig:theory}(c)].

\textit{Conclusion.} We presented confocal PLE spectroscopy as a method to measure a variety of optical excitations in Cu$_{2}$O: yellow and green Rydberg excitons, blue excitons, yellow and green continuum and Urbach states, as well as phonon-assisted absorption into excitonic states. A quantitative analysis of the spectroscopic method is given. Since the optical emission from the yellow $1s$ orthoexcitons was detected, the observed PLE spectrum reveals information about the transition probabilities from the various excitations to the yellow $1s$ orthoexciton state. A model for these transition probabilities is presented, which can be used as input into theoretical calculations to obtain a deeper understanding of the excitonic transitions and interactions in Cu$_{2}$O. By PLE spectroscopy on a well-polished and etched crystal in a dilution refrigerator, we have been able to identify giant yellow Rydberg $p$ excitons up to $n=30$, corresponding to a diameter of 3.0 $\mu$m and a blockade diameter of 23 $\mu$m \cite{kazimierczuk2014}. The observation of these high Rydberg states is important, since all nonlinearities scale drastically with $n$, and therefore the discovery of every new Rydberg state increases possibilities for applications in quantum technology \cite{walthernatcomm2018}. Interestingly, a recent theoretical paper predicts dominance of radiative transitions and appearance of polaritonic features for $n>28$ \cite{stolz2018}, which opens up exciting opportunities for future studies.

We thank D. Fr\"ohlich for his advice on sample preparation. We acknowledge support from the Swedish Research Council (Starting Grants 2016-04527 and 2019-04821) and from the Göran Gustafsson Foundation.


\begin{thebibliography}{99}
\bibitem{adams2020}
C. S. Adams, J. D. Pritchard, and J. P. Shaffer, "Rydberg atom quantum technologies," J. Phys. B: At. Mol. Opt. Phys. \textbf{53,} 012002 (2020). 
\bibitem{assmann2020}
M. Assmann and M. Bayer, "Semiconductor Rydberg Physics," Adv. Quantum Technol. \textbf{3,} 1900134 (2020). 
\bibitem{waltherprb2018}
V. Walther, S. O. Kr\"uger, S. Scheel, and T. Pohl, "Interactions between Rydberg excitons in Cu$_2$O," Phys. Rev. B \textbf{98,} 165201 (2018). 
\bibitem{walthernatcomm2018}
V. Walther, R. Johne, and T. Pohl, "Giant optical nonlinearities from Rydberg excitons in semiconductor microcavities," Nat. Comm. \textbf{9,} 1309 (2018).
\bibitem{gross1956}
E. F. Gross, "Optical Spectrum of Excitons in the Crystal Lattice," Nuovo Cimento \textbf{3,} 672 (1956). 
\bibitem{agekyan1977}
V. T. Agekyan, "Spectroscopic properties of semiconductor crystals with direct forbidden energy gap," Phys. Stat. Solidi (a) \textbf{43,} 11 (1977). 
\bibitem{chernikov2014}
A. Chernikov,  T. C. Berkelbach, H. M. Hill, A. Rigosi, Y. Li, O. Burak Aslan, D. R. Reichman, M. S. Hybertsen, and T. F. Heinz, "Exciton binding energy and nonhydrogenic Rydberg series in monolayer WS$_2$," Phys. Rev. Lett. \textbf{113,} 076802 (2014). 
\bibitem{he2014}
K. He, N. Kumar, L. Zhao, Z. Wang, K. Fai Mak, H. Zhao, and J. Shan, "Tightly bound excitons in monolayer WSe$_2$," Phys. Rev. Lett. \textbf{113,} 026803 (2014). 
\bibitem{hill2015}
H. M. Hill, A. F. Rigosi, C. Roquelet, A. Chernikov, T. C. Berkelbach, D. R. Reichman, M. S. Hybertsen, L. E. Brus, and T. F. Heinz, "Observation of excitonic Rydberg states in monolayer MoS$_2$ and WS$_2$ by photoluminescence excitation spectroscopy," Nano Lett. \textbf{15,} 2992 (2015). 
\bibitem{wang2020}
T. Wang, Z. Li, Y. Li, Z. Lu, S. Miao, Z. Lian, Y. Meng, M. Blei, T. Taniguchi, K. Watanabe, S. Tongay, D. Smirnov, C. Zhang, and S. F. Shi, "Giant Valley-Polarized Rydberg Exciton in Monolayer WSe$_2$ Revealed by Magneto-photocurrent Spectroscopy," Nano Lett. \textbf{20,} 7635 (2020). 
\bibitem{kazimierczuk2014}
T. Kazimierczuk, D. Fr\"ohlich, S. Scheel, H. Stolz, and M. Bayer, "Giant Rydberg excitons in the copper oxide Cu$_2$O," Nature \textbf{514,} 343 (2014). 
\bibitem{heckotter2020}
J. Heck\"otter, D. Janas, R. Schwartz, M. Assmann, and M. Bayer, "Experimental limitation in extending the exciton series in Cu$_2$O towards higher principal quantum numbers," Phys. Rev. B \textbf{101,} 235207 (2020). 
\bibitem{heckotterprl2018}
J. Heck\"otter, M. Freitag, D. Fr\"ohlich, M. Assmann, M. Bayer, P. Gr\"unwald, F. Sch\"one, D. Semkat, H. Stolz, and S. Scheel, "Rydberg Excitons in the Presence of an Ultralow-Density Electron-Hole Plasma," Phys. Rev. Lett. \textbf{121,} 097401 (2018). 
\bibitem{walther2020}
V. Walther and T. Pohl, "Plasma-Enhanced Interaction and Optical Nonlinearities of Cu$_2$O Rydberg Excitons," Phys. Rev. Lett. \textbf{125,} 097401 (2020). 
\bibitem{mysyrowicz1979}
A. Mysyrowicz, D. Hulin, and A. Antonetti, "Long Exciton Lifetime in Cu$_2$O," Phys. Rev. Lett. \textbf{43,} 1123 (1979). 
\bibitem{snoke2014}
D. Snoke and G. M. Kavoulakis, "Bose–Einstein condensation of excitons in Cu$_2$O: progress over 30 years," Rep. Prog. Phys. \textbf{77,} 116501 (2014). 
\bibitem{heckotterefield2017}
J. Heck\"otter, M. Freitag, D. Fr\"ohlich, M. Assmann, M. Bayer, M. A. Semina, and M. M. Glazov, "High-resolution study of the yellow excitons in Cu$_2$O subject to an electric field," Phys. Rev. B \textbf{95,}  035210 (2017). 
\bibitem{heckotterprb2018}
J. Heck\"otter, M. Freitag, D. Fr\"ohlich, M. Assmann, M. Bayer, M. A. Semina, and M. M. Glazov, "Dissociation of excitons in Cu$_2$O by an electric ﬁeld," Phys. Rev. B \textbf{98,} 035150 (2018). 
\bibitem{assmann2016}
M. Assmann, J. Thewes, D. Fr\"ohlich, and M. Bayer, "Quantum chaos and breaking of all antiunitary symmetries in Rydberg excitons," Nat. Mater. \textbf{15,} 741 (2016). 
\bibitem{heckotterscalinglaws2017}
J. Heck\"otter, M. Freitag, D. Fr\"ohlich, M. Assmann, M. Bayer, M. A. Semina, and M. M. Glazov, "Scaling laws of Rydberg excitons," Phys. Rev. B \textbf{96,} 125142 (2017). 
\bibitem{zielinska2017}
S. Zielinska-Raczynska, D. Ziemkiewicz, and G. Czajkowski, "Magneto-optical properties of Rydberg excitons: Center-of-mass quantization approach," Phys. Rev. B \textbf{95,} 075204 (2017). 
\bibitem{artyukhin2018}
S. Artyukhin , D. Fishman, C. Faugeras, M. Potemski, A. Revcolevschi, M. Mostovoy, and P. H. M. van Loosdrecht, "Magneto-absorption spectra of hydrogen-like yellow exciton series in cuprous oxide: excitons in strong magnetic fields," Sci. Rep. \textbf{8,} 7818 (2018). 
\bibitem{farenbruch2020}
A. Farenbruch, D. Fr\"ohlich, D. R. Yakovlev, and M. Bayer, "Rydberg Series of Dark Excitons in Cu$_2$O," Phys. Rev. Lett. \textbf{125,} 207402 (2020).
\bibitem{grunwald2016} 
P. Gr\"unwald, M. Assmann, J. Heck\"otter, D. Fr\"ohlich, M. Bayer, H. Stolz, and S. Scheel, "Signatures of Quantum Coherences in Rydberg Excitons," Phys. Rev. Lett. \textbf{117,} 133003 (2016). 
\bibitem{schweiner2016}
F. Schweiner, J. Main, and G. Wunner, "Linewidths in excitonic absorption spectra of cuprous oxide," Phys. Rev. B \textbf{93,} 085203 (2016). 
\bibitem{schone2017}
F. Sch\"one, H. Stolz, and N. Naka, "Phonon-assisted absorption of excitons in Cu$_2$O," Phys. Rev. B \textbf{96,} 115207 (2017). 
\bibitem{stolz2018}
H. Stolz, F. Sch\"one, and D. Semkat, "Interaction of Rydberg excitons in cuprous oxide with phonons and photons: optical linewidth and polariton effect," New J. Phys. \textbf{20,} 023019 (2018). 
\bibitem{kruger2020}
S. O. Kr\"uger, H. Stolz, and S. Scheel, "Interaction of charged impurities and Rydberg excitons in cuprous oxide," Phys. Rev. B \textbf{101,} 235204 (2020). 
\bibitem{takahata2018}
M. Takahata and N. Naka, "Photoluminescence properties of the entire excitonic series in Cu$_2$O," Phys. Rev. B  \textbf{98,} 195205 (2018). 
\bibitem{kitamura2017}
T. Kitamura, M. Takahata, and N. Naka, "Quantum number dependence of the photoluminescence broadening of excitonic Rydberg states in cuprous oxide," J. Lumin. \textbf{192,} 808 (2017). 
\bibitem{yu1978}
P. Y. Yu and Y. R. Shen, "Resonance Raman studies in Cu$_2$O. II. The yellow and green excitonic series," Phys. Rev. B \textbf{17,} 4017 (1978). 
\bibitem{jorger2003}
E. T. E. J\"orger, T. Fleck, and C. Klingshirn, "Infrared absorption by excitons in Cu$_2$O," Phys. stat. sol. (b) \textbf{238,} 470 (2003).
\bibitem{yoshioka2007} 
K. Yoshioka, T. Ideguchi, and M. Kuwata-Gonokami, "Laser-based continuous-wave excitonic Lyman spectroscopy in Cu$_2$O," Phys. Rev. B \textbf{76,} 033204 (2007). 
\bibitem{takahatanonlocal2018}
M. Takahata, K. Tanaka, and N. Naka, "Nonlocal optical response of weakly conﬁned excitons in Cu$2$O mesoscopic ﬁlms," Phys. Rev. B \textbf{97,} 205305 (2018). 
\bibitem{rommel2020}
P. Rommel, P. Zielinski, and J. Main, "Green exciton series in cuprous oxide," Phys. Rev. B \textbf{101,} 075208 (2020). 
\bibitem{kruger2019}
S. O. Kr\"uger and S. Scheel, "Interseries transitions between Rydberg excitons in Cu$_2$O," Phys. Rev. B \textbf{100,} 085201 (2019). 
\bibitem{schmultzler2013}
J. Schmutzler, D. Fr\"ohlich, and M. Bayer, "Signatures of coherent propagation of blue polaritons in Cu$_2$O," Phys. Rev. B \textbf{87,} 245202 (2013). 
\bibitem{ziemkiewicz2020}
D. Ziemkiewicz, "Electromagnetically Induced Transparency in Media with Rydberg Excitons 1: Slow Light," Entropy \textbf{22,} 177 (2020). 
\bibitem{walthereit2020}
V. Walther, P. Gr\"unwald, and T. Pohl, "Controlling Exciton-Phonon Interactions via Electromagnetically Induced Transparency," Phys. Rev. Lett. \textbf{125,} 173601 (2020). 
\bibitem{steinhauer2020}
S. Steinhauer, M. A. M. Versteegh, S. Gyger, A. W. Elshaari, B. Kunert, A. Mysyrowicz, and V. Zwiller, "Rydberg excitons in Cu$_{2}$O microcrystals grown on a silicon platform," Comm. Mater. \textbf{1,} 11 (2020). 
\bibitem{konzelmann2020}
A. Konzelmann, B. Frank, and H. Giessen, "Quantum conﬁned Rydberg excitons in reduced dimensions," J. Phys. B: At. Mol. Opt. Phys. \textbf{53,} 024001 (2020). 
\bibitem{khazali2017}
M. Khazali, K. Heshami, and C. Simon, "Single-photon source based on Rydberg exciton blockade," J. Phys. B: At. Mol. Opt. Phys. \textbf{50,} 215301 (2017). 
\bibitem{snoke1987}
D. Snoke, J. P. Wolfe, and A. Mysyrowicz, "Quantum Saturation of a Bose Gas: Excitons in Cu$_2$O," Phys. Rev. Lett. \textbf{59,} 827 (1987). 
\bibitem{snoke1990}
D. W. Snoke, J. P. Wolfe, and A. Mysyrowicz, "Evidence for Bose-Einstein Condensation of a Two-Component Exciton Gas," Phys. Rev. Lett. \textbf{64,} 2543 (1990). 
\bibitem{supplementalmaterial}
Details of the sample preparation, power stabilization, and calculations of the absorption spectrum and the detection probability as a function of exciton position and a discussion of the model for the relaxation to the $1s$ orthoexciton state are given as Supplemental Material.
\bibitem{yu1975}
P. Y. Yu and Y. R. Shen, "Resonance Raman studies in Cu$_2$O. I. The phonon-assisted $1s$ yellow excitonic absorption edge," Phys. Rev. B \textbf{12,} 1377 (1975). 
\bibitem{toyozawa1964}
Y. Toyozawa, "Interband effect of lattice vibrations in the exciton absorption spectra," J. Phys. Chem. Solids . \textbf{25,} 59 (1964). 
\bibitem{trauernicht1986}
D. P. Trauernicht and J. P. Wolfe, "Drift and diffusion of paraexcitons in Cu$_2$O: Deformation-potential scattering in the low-temperature regime," Phys. Rev. B \textbf{33,} 8506 (1986). 
\bibitem{morita2019}
Y. Morita, H. Suzuki, K. Yoshioka, and M. Kuwata-Gonokami, "Observation of ultrahigh mobility excitons in a strain field by space- and time-resolved spectroscopy at subkelvin temperatures," Phys. Rev. B \textbf{100,} 035206 (2019). 
\bibitem{yoshioka2010}
K. Yoshioka, T. Ideguchi, A. Mysyrowicz, and M. Kuwata-Gonokami, "Quantum inelastic collisions between paraexcitons in Cu$_2$O," Phys. Rev. B \textbf{82,} 041201(R) (2010). 
\bibitem{webb1996}
R. H. Webb, "Confocal optical microscopy," Rep. Prog. Phys. \textbf{59,} 427 (1996). 
\bibitem{ahi2017}
K. Ahi, "Mathematical Modeling of THz Point Spread Function and Simulation of THz Imaging Systems," IEEE Trans. THz Sci. Technol. \textbf{7,} 747 (2017). 
\bibitem{schweiner2017}
F. Schweiner, J. Main, G. Wunner, and C. Uihlein, "Even exciton series in Cu$_2$O," Phys. Rev. B \textbf{95,} 195201 (2017). 
\bibitem{daunois1966}
A. Daunois, J. L. Deiss, and B. Meyer, "\'Etude spectrophotom\'etrique de l'absorption bleue et violette de Cu$_2$O," J. Phys. \textbf{27,} 142 (1966).
\end{thebibliography}
\end{document}


\title{Giant Rydberg excitons in Cu$_{2}$O probed by photoluminescence excitation spectroscopy: Supplemental Material}

\author{Marijn A. M. Versteegh}
\email{verst@kth.se}
 \affiliation{Department of Applied Physics, KTH Royal Institute of Technology, Stockholm, Sweden}
\author{Stephan Steinhauer}
\affiliation{Department of Applied Physics, KTH Royal Institute of Technology, Stockholm, Sweden}
\author{Josip Bajo}
\affiliation{Department of Applied Physics, KTH Royal Institute of Technology, Stockholm, Sweden}
\author{Thomas Lettner}
\affiliation{Department of Applied Physics, KTH Royal Institute of Technology, Stockholm, Sweden}
\author{Ariadna Soro}
\altaffiliation[Present address: ]{Department of Microtechnology and Nanoscience, Chalmers University of Technology, Gothenburg, Sweden}
\affiliation{Department of Applied Physics, KTH Royal Institute of Technology, Stockholm, Sweden}
\author{Alena Romanova}
\affiliation{Department of Applied Physics, KTH Royal Institute of Technology, Stockholm, Sweden}
\affiliation{Johannes Gutenberg University of Mainz, Germany}
\author{Samuel Gyger}
\affiliation{Department of Applied Physics, KTH Royal Institute of Technology, Stockholm, Sweden}
\author{Lucas Schweickert}
\affiliation{Department of Applied Physics, KTH Royal Institute of Technology, Stockholm, Sweden}
\author{Andr\'e Mysyrowicz}
\affiliation{Laboratoire d’Optique Appliqu\'ee, ENSTA, CNRS, \'Ecole Polytechnique, Palaiseau, France}
\author{Val Zwiller}
\affiliation{Department of Applied Physics, KTH Royal Institute of Technology, Stockholm, Sweden}

\date{\today}

\maketitle

\section{Sample preparation}
Before measurement, the Cu$_2$O crystals were carefully polished using MicroCloth and an alkaline colloidal silica polishing solution (Ludox), until a mirror-like surface was obtained and even $\mu$m-sized dips at the surface had disappeared. After polishing, the crystals were etched using hydrochloric acid (HCl) and ammonia (NH$_3$) \cite{takahata2018}. First, a polished crystal was held for 10 s in the HCl solution (2 mol/l), then rinsed with distilled water and blown dry with N$_2$. Second, the crystal was held for 10 s in ammonia (13 mol/l), then rinsed with water and blown dry with N$_2$. At this point, the crystal had an opaque appearance with blue/grey color. After carefully removing loose material from the surface with a wet foam head swap (Texwipe), the crystal retained its transparency. This etching procedure was repeated two more times. 
The quality of surface preparation was crucial for the appearance of giant Rydberg excitons. Polishing with the colloidal silica solution gave the best polishing result, and this, in turn, resulted in the observation of the highest Rydberg states. Surface charges have a detrimental effect on giant Rydberg excitons, because of the Stark effect \cite{kruger2020, heckotter2020}. In order to minimize oxidation, we mounted the cuprite crystal directly after etching in the dilution refrigerator, which was immediately evacuated. In vacuum, the crystal quality can be maintained at least for several months. We have observed that when the crystal is kept in air for several days, however, defect emission appears in the PL spectrum and the highest Rydberg excitons are no longer observed. The original spectrum can then be recovered by careful polishing and etching of the surface. 

\renewcommand{\thefigure}{S\arabic{figure}}
\begin{figure}
\includegraphics[width=0.47\textwidth]{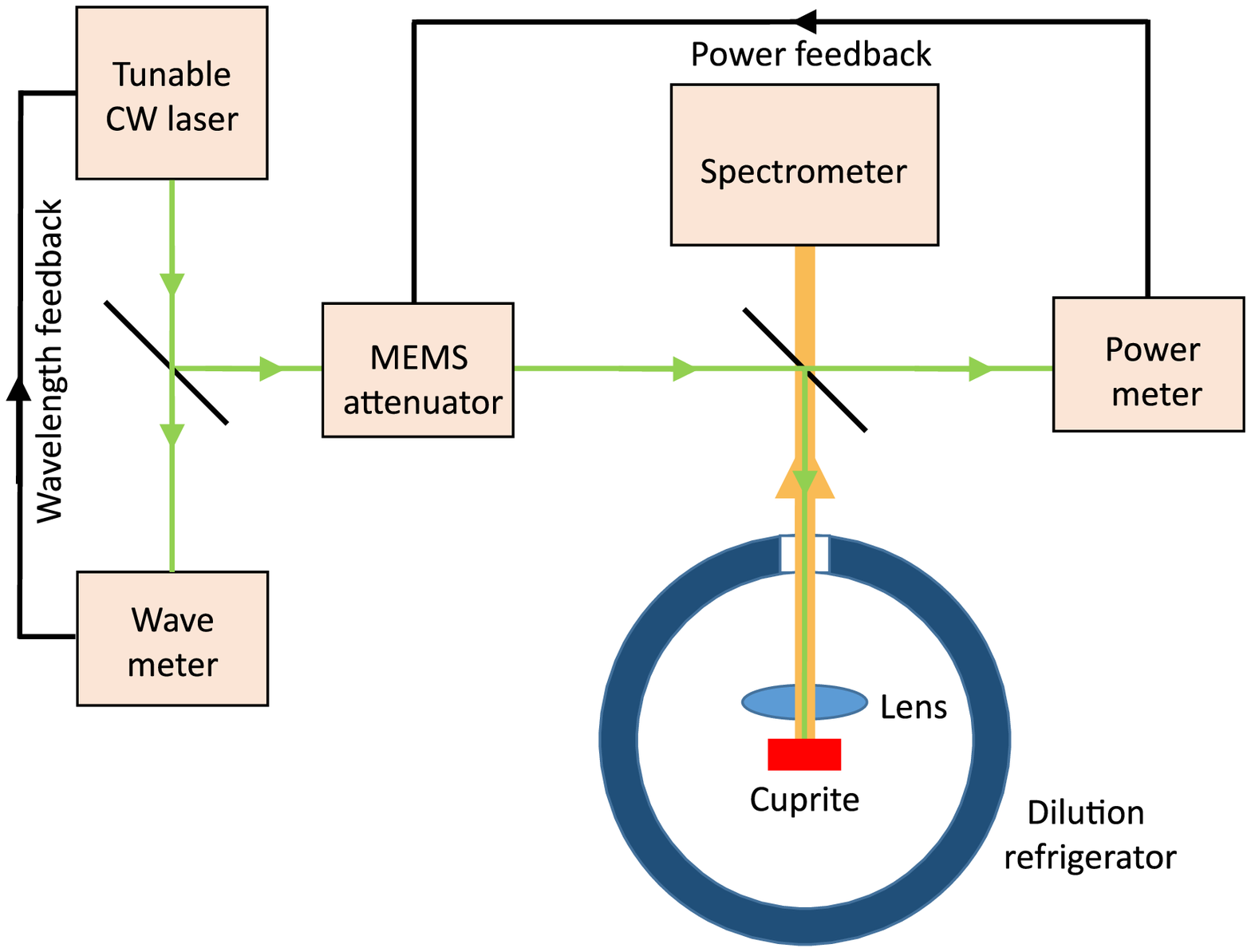}
\caption{\label{fig:setup} Schematic of the experimental setup.}
\end{figure}

\section{Power stabilization}
A schematic of the setup is given in Fig. \ref{fig:setup}. The input power of the excitation laser is attenuated using a MEMS based variable attenuator (Thorlabs, V450A).
In front of the dilution refrigerator a plate beam splitter is placed, which reflects 10\% of the laser power into the dilution refrigerator and transmits 90\% of the optical signal from the Cu$_2$O crystal towards the spectrometer. The 90\% of the laser power that is transmitted by this plate beam splitter is sent to a power meter (Thorlabs PM100USB with S120C Powerhead). The splitting ratio is calibrated using a handheld power meter. The measured power is used as set point in a software controlled PID controller (LabView).
The generated control signal is sent to a homebuild Digital Analog Converter based on an DAC-IC (AD5541A, Analog Devices) controlled by an ESP32 Feather board (Adafruit). With this setup, the power fluctuations are smaller than 0.1\%.

\section{Absorption spectrum}
The yellow Rydberg $np$ resonances are modeled by asymmetric Lorentzians \cite{toyozawa1964, kruger2020}
\renewcommand{\theequation}{S\arabic{equation}}	
\begin{equation}
\alpha_{np,y}(\omega)=C_{np,y}\frac{\Gamma_{np,y}/2+2\xi_{n,y}g\left(\frac{\hbar(\omega-\omega_n)}{\Gamma_{np,y}/2}\right)\hbar(\omega-\omega_n)}{(\Gamma_{np,y}/2)^2+\hbar^2(\omega-\omega_n)^2},
\end{equation}
where $C_{np,y}$ describes the oscillator strength, $\Gamma_{np,y}$ the linewidth (full width at half maximum), $\omega_n$ the resonance frequency, $\xi_{n,y}$ the asymmetry parameter, and $g(b)$ is a function included to reproduce long-range decay of symmetric Lorentzians \cite{kruger2020} and is defined as
\begin{equation}
 g(b)= \left\{
 \begin{array}{ll}
	 1 & \textrm{if } |b|<6;\\
	e^{-\left(\frac{|b|-6}{6}\right)^2} & \textrm{else.}
 \end{array} \right.
\end{equation}
We find that the parameter 6 yields an optimal fit in describing the transition between the asymmetric and the symmetric Lorentzian lineshape at $3\Gamma_{np,y}$.

The absorption spectrum and the contribution of the various components to the absorption spectrum are presented in Fig. 3(a) in the main article. The constants and fitting parameters relevant to the absorption spectrum are given in Tables S1 and S2. The obtained oscillator strengths $C_{np,y}$ are given in Fig. 2(e) and the asymmetry parameters $\xi_{n,y}$ are given in Fig. 2(f) in the main article.

\renewcommand{\thetable}{S\arabic{table}}																																		
\begin{table}
\caption{Constants for the calculation of the absorption spectrum.}
\begin{ruledtabular}
\begin{tabular}{ll}
\textbf{Constant} & \textbf{Value}\\
\hline
Normalized yellow band gap & $E_{\textrm{gap},y}= 2.17196$ eV\\
& (determined via PLE)\\
Green band gap & $E_{\textrm{gap},g}= 2.30302$ eV \cite{rommel2020}\\
$1s$ yellow orthoexciton energy & $E_{1s,\textrm{ortho}} = 2.03351$ eV\\
& (determined via PL)\\
Yellow Rydberg constant & Ry$_{y}= 86$ meV \cite{heckotterprl2018}\\
$\Gamma_{3}^{-}$ phonon energy & $\hbar\omega_{\Gamma_{3}^{-}}= 13.6$ meV \cite{yu1975}\\
$\Gamma_{4}^{-}$ phonon energy & $\hbar\omega_{\Gamma_{4}^{-}}= 82.1$ meV \cite{yu1978}\\
$1s$ exciton mass & $2.61 m_0$ \cite{brandt2007}\\
Yellow Bohr radius & 0.81 nm \cite{schweiner2017}\\
Blue Bohr radius & 1.72 nm \cite{schone2017}\\
$1s$ blue exciton energy & 2.569 eV \cite{schone2017}\\
$2s$ blue exciton energy & 2.611 eV \cite{schone2017}\\
Quadratic deformation parameter & 0.168 nm$^2$ \cite{schone2017}\\
\end{tabular}\end{ruledtabular}
\end{table}

\begin{table}
\caption{Fit parameters for the calculation of the absorption spectrum.}
\begin{ruledtabular}
\begin{tabular}{ll}
\textbf{Fit parameter} & \textbf{Value}\\
\hline
Nominal yellow band gap & $E_{\textrm{gap0},y}= 2.172053$ eV\\
& (agreement with \cite{heckotterprl2018})\\
Point defect in the yellow series & $\delta_P= 0.21$\\
Minimum yellow Rydberg linewidth & $\Gamma_0=5.55$ $\mu$eV at 2 nW,\\
& $\Gamma_0=7$ $\mu$eV at 10 nW\\ 
Yellow Rydberg linewidth parameter & $\Gamma_1= 19$ meV\\
Yellow Rydberg oscillator strengths & $C_{np,y}$, see Fig. 2(e)\\
Rydberg asymmetry parameters & $\xi_{n,y}$, see Fig. 2(f)\\
$1s$ green exciton energy & $E_{1s,g} = 2.148$ eV\\
$2p$ green exciton energy & $E_{2p,g} = 2.269$ eV\\
$3p$ green exciton energy & $E_{3p,g} = 2.290$ eV\\
Yellow and green Urbach parameter & $E_U = 20$ meV\\
\end{tabular}\end{ruledtabular}
\end{table}

\section{Detection probability}
In this section, we calculate the relative detection probability for a photon emitted by a $1s$ orthoexciton in the Cu$_2$O crystal as a function of the exciton position $(x,y,z)$. Here, the $xy$ plane is the front surface of the crystal, $z$ is the depth and the focus of the lens is at $(0,0,0)$. For a photon to be detected, it needs to be transmitted at the crystal-air interface, collected by the lens ($f = 3.1$ mm, NA = 0.70), and enter the spectrometer. A photon emitted from an exciton at the focus has the highest probability of being detected, since a wide solid angle is available for the photon to be captured, in contrast to a photon emitted deeper in the crystal. Refraction and total internal reflection at the crystal surface have to be taken into account, reducing the detection angle further. 
Following approaches in literature \cite{webb1996, ahi2017}, we write the probability of detecting a photon emitted at position $(x,y,z)$ as
\begin{multline}
P_{\textrm{det}}(x,y,z)=\\
\frac{2c_1}{\pi [w_{\textrm{det}}(z)]^2}\int k(t,u,z)e^{\frac{-2(x-t)^2-2(y-u)^2}{[w_{\textrm{det}}(z)]^2}}\textrm{d}t\textrm{d}u,
\end{multline}
where $c_1$ 
is a normalization constant, $w_{\textrm{det}}(z)=w_{\textrm{det},0}\sqrt{1+(z/z_{R,\textrm{det}})^2}$ 
is the detection waist at depth $z$ and $z_{R,\textrm{det}}=\pi w_{\textrm{det},0}^2 n_{\textrm{det}}/\lambda_{\textrm{det}}$ is the detection Rayleigh range, with $n_{\textrm{det}}=2.988$ the index of refraction at detection wavelength $\lambda_{\textrm{det}}=610$ nm, and $w_{\textrm{det},0}=\lambda_{\textrm{det}}/\{\pi\tan[\arcsin(\textrm{NA})]\} = 0.20$ $\mu$m the detection waist at the focus. The function $k(t,u,z)$ is given by
\begin{equation}
k(t,u,z)=\left\{
\begin{array}{ll}
\frac{1}{\pi [r_1(z)]^2} & \textrm{if } t^2+u^2\leq [r_1(z)]^2\\
0 & \textrm{else},
 \end{array} \right.
\end{equation}  
with $r_1 (z)=r_0+z r_0/(f n_{\textrm{det}})\approx r_0 = 7.3$ $\mu$m the geometrical radius of the detection area on the crystal as determined by the finite aperture stop of the detection system. The optical absorption at 610 nm in the Cu$_2$O crystal is negligible. The calculated relative detection probability $P_\textrm{det}(x,y,z)$ is shown in the main text in Fig. 3(b), for the case of $y=0$. 

\section{Relaxation model}
By fitting the theoretical curve for the PLE signal to the experimental data we arrive at functions for the probabilities of transitions from various excited states to the yellow $1s$ orthoexciton state, given by Eqs. (6)-(13) in the main text. In this section, we discuss a qualitative physical interpretation of these transition probabilities, first the relaxation from the yellow states and then the relaxation from the green states to the yellow $1s$ orthoexciton state.

For the relaxation of yellow $1s$ orthoexcitons, created by $\Gamma_{3}^{-}$ or $\Gamma_{4}^{-}$ phonon-assisted absorption to the thermalized $1s$ orthoexciton state (from which the emission is detected) we find
\begin{equation}\label{yellow1sphononassisted}
P_{1s,y}^{\Gamma_{i}^{-}}(\omega)=e^{-(\hbar\omega-E_{1s,y,\textrm{ortho}}-\hbar\omega_{\Gamma_{i}^{-}})/E_{\phi}},
\end{equation}
where $i=3,4$. For $1s$ orthoexcitons created with minimal kinetic energy, that is with photon energies $\hbar\omega$ only slightly above $E_{1s,y,\textrm{ortho}}+\hbar\omega_{\Gamma_{i}^{-}}$, the probability of relaxation to thermalized $1s$ orthoexcitons is unity by definition. This is how the overall fitting factor was determined. We find a good fit with relaxation probability that decays exponentially, which can be understood from the fact that for highly energetic orthoexcitons the cooling process involves emission of many phonons, with a larger probability of collisions leading to dissociation or nonradiative recombination. The fit parameter here is $E_\phi$.

For the yellow Rydberg excitons (from $2p$ to $35p$) we have
\begin{equation}
P_{np,y}(\omega)=\frac{3}{4}.
\end{equation}
Yellow $np$ Rydberg excitons relax with high probability to the yellow $1s$ states via optical phonon emission \cite{toyozawa1964}. The factor $\frac{3}{4}$ comes from the fact that the $1s$ orthoexciton state is a triplet state, while the $1s$ paraexciton state is a singlet state, so that three quarters of the decaying yellow Rydberg excitons arrive at an orthoexciton state.

For the continuum states above the yellow band gap 
\begin{equation}
P_{p,y,\textrm{cont}}(\omega)=\frac{3}{4}de^{-[(\hbar\omega-E_{\textrm{gap},y})/E_{i}]^2}.
\end{equation}
Here, $d$ and $E_i$ are fit parameters. Yellow continuum states (uncorrelated electron-hole pairs) scatter via phonon emission into yellow $np$ Rydberg states, before relaxing further into the yellow $1s$ states \cite{yu1978}. This transition probability can therefore be regarded as the product of the transition probability from a yellow continuum state to a yellow $np$ Rydberg state, in this model given by $de^{-[(\hbar\omega-E_{\textrm{gap},y})/E_{i}]^2}$, and the transition probability from a yellow $np$ Rydberg state to the $1s$ orthoexciton state, which is $\frac{3}{4}$. The factor $d<1$ describes the fact that photon absorption creates an uncorrelated electron-hole pair, not a Coulomb-bound exciton. Transition to a $1s$ orthoexciton requires formation of an exciton. For higher excitations into the bands the transition probability decreases stronger than exponentially, because the decay process involves a large number of phonon emissions and, in addition, uncorrelated electrons and holes with a high kinetic energy have a large probability of separating from each other and never form an excitonic state.

For the states in the yellow Urbach tail
\begin{equation}
P_{p,y,\textrm{Urbach}}(\omega)=\frac{3}{4}d.
\end{equation}
States in the Urbach tail can also decay via phonon emission to yellow $np$ Rydberg states, which decay further to the $1s$ orthoexciton state with probability $3/4$. Continuity at the band gap requires the inclusion of the same fit parameter $d$.

For the $\Gamma_{3}^{-}$ and $\Gamma_{4}^{-}$ phonon-assisted absorption into the $1s$ green excitons
\begin{equation}
P_{1s,g}^{\Gamma_{i}^{-}}(\omega)=\frac{3}{4}de^{-(\hbar\omega-E_{1s,g}-\hbar\omega_{\Gamma_{i}^{-}})/E_{\phi}},
\end{equation}
where $i=3,4$. Here, we have the same exponential as in Eq. (\ref{yellow1sphononassisted}), describing the probability of relaxation to a low kinetic energy. Green $1s$ excitons strongly couple to the yellow $2p$ state \cite{schweiner2017}. The parameter $d$ here describes the transition probability from the low-energy $1s$ green state to the yellow $2p$ state and is considered in this model to be equal to the transition probability from a low-energy yellow continuum state, or a yellow Urbach state. The relaxation from the yellow $2p$ state to the yellow $1s$ orthoexciton state again has probability $3/4$. The energy of the $1s$ green exciton, $E_{1s,g}$, is used as a fit parameter and we obtain the best fit with $E_{1s,g} = 2.148$ eV, which is equal to $E_{2p,y}$. However, due to the strong coupling with the yellow $2p$ state, the $1s$ green exciton resonance is  spread over several resonances \cite{schweiner2017}.

For the green Rydberg excitons ($2p$ and $3p$)
\begin{equation}
P_{np,g}(\omega)=\frac{3}{4}h.
\end{equation}
Green Rydberg excitons make a transition to a yellow continuum state, followed by decay to yellow Rydberg states and finally to the yellow $1s$ state \cite{yu1978}. The probabilities of the transitions to the yellow Rydberg states are brought together in the single fit parameter $h$. Because of the complicated pathway, involving the transition of a hole from the $\Gamma_8^+$ to the $\Gamma_7^+$ band, $h$ should be much smaller than $d$.

For the continuum states above the green band gap the transition to the $1s$ orthoexciton is even more unlikely, as first the transition is to be made to one of the green Rydberg $p$ states \cite{yu1978}. Similar to the transitions from the yellow continuum, we take as transition probability to the Rydberg states a factor $de^{-[(\hbar\omega-E_{\textrm{gap},g})/E_{i}]^2}$, resulting in a total probability of transition from a green continuum state to the yellow $1s$ orthoexciton state of
\begin{equation}
P_{p,g,\textrm{cont}}(\omega)=\frac{3}{4}hde^{-[(\hbar\omega-E_{\textrm{gap},g})/E_{i}]^2}.
\end{equation}
No new fit parameter is introduced in our model here.

For the states in the green Urbach tail, continuity at the green band gap requires that 
\begin{equation}
P_{p,g,\textrm{Urbach}}(\omega)=\frac{3}{4}hd.
\end{equation}

The obtained values for the fit parameters are $d= 0.687$, $h= 0.187$, $E_i= 187$ meV and $E_\phi= 0.28$ eV. The full range of the fit is presented in the main text as the black curve in Fig. 2(a). The same fit is used in Figs. 2(c) and 2(d), with oscillator strengths and asymmetry parameters as given in Figs. 2(e) and 2(f).